\documentclass[10pt,conference]{IEEEtran}
\usepackage{cite,calc}
\usepackage[cmex10]{amsmath}   
\usepackage{amsfonts,amssymb,graphics,psfrag,subfigure,epsfig,graphics}
\usepackage{epstopdf}
\usepackage[mathcal]{euscript}                 
\usepackage{multirow}
\usepackage{colortbl}
\usepackage{arydshln}

\newcommand{\blath}[1]{\mbox{#1-th}}

\newcommand{\old}[1]{}
\newcommand{\rem}[1]{}

\newcommand{\secref}[1]{Section~\ref{#1}}
\newcommand{\thrmref}[1]{Theorem~\mbox{\ref{#1}}}
\newcommand{\appref}[1]{Appendix~\mbox{\ref{#1}}}
\newcommand{\lemref}[1]{Lemma~\ref{#1}}

\newcommand{\JET}{\mathrm{JET}}
\newcommand{\GMD}{\mathrm{GMD}}

\newtheorem{remark}{Remark}
\newtheorem{thm}{Theorem}

\newtheorem{lemma}{Lemma}
\newtheorem{example}{Example}

\newcommand{\trace}[1]{\mathrm{trace}\left( #1 \right)}

\newcommand{\bX}{{\bf X}}
\newcommand{\bx}{{\bf x}}
\newcommand{\bY}{{\bf Y}}
\newcommand{\by}{{\bf y}}
\newcommand{\bZ}{{\bf Z}}

\newcommand{\bz}{{\bf z}}

\DeclareMathAlphabet{\mathbsf}{OT1}{cmss}{bx}{n}
\DeclareMathAlphabet{\mathssf}{OT1}{cmss}{m}{sl}
\DeclareMathAlphabet{\mathcsf}{OT1}{cmss}{sbc}{n}

\DeclareSymbolFont{bsfletters}{OT1}{cmss}{bx}{n}  
\DeclareSymbolFont{ssfletters}{OT1}{cmss}{m}{n}
\DeclareMathSymbol{\bsfGamma}{0}{bsfletters}{'000}
\DeclareMathSymbol{\ssfGamma}{0}{ssfletters}{'000}
\DeclareMathSymbol{\bsfDelta}{0}{bsfletters}{'001}
\DeclareMathSymbol{\ssfDelta}{0}{ssfletters}{'001}
\DeclareMathSymbol{\bsfTheta}{0}{bsfletters}{'002}
\DeclareMathSymbol{\ssfTheta}{0}{ssfletters}{'002}
\DeclareMathSymbol{\bsfLambda}{0}{bsfletters}{'003}
\DeclareMathSymbol{\ssfLambda}{0}{ssfletters}{'003}
\DeclareMathSymbol{\bsfXi}{0}{bsfletters}{'004}
\DeclareMathSymbol{\ssfXi}{0}{ssfletters}{'004}
\DeclareMathSymbol{\bsfPi}{0}{bsfletters}{'005}
\DeclareMathSymbol{\ssfPi}{0}{ssfletters}{'005}
\DeclareMathSymbol{\bsfSigma}{0}{bsfletters}{'006}
\DeclareMathSymbol{\ssfSigma}{0}{ssfletters}{'006}
\DeclareMathSymbol{\bsfUpsilon}{0}{bsfletters}{'007}
\DeclareMathSymbol{\ssfUpsilon}{0}{ssfletters}{'007}
\DeclareMathSymbol{\bsfPhi}{0}{bsfletters}{'010}
\DeclareMathSymbol{\ssfPhi}{0}{ssfletters}{'010}
\DeclareMathSymbol{\bsfPsi}{0}{bsfletters}{'011}
\DeclareMathSymbol{\ssfPsi}{0}{ssfletters}{'011}
\DeclareMathSymbol{\bsfOmega}{0}{bsfletters}{'012}
\DeclareMathSymbol{\ssfOmega}{0}{ssfletters}{'012}

\newcommand{\Capacity}[1]{I(H_{#1},C_\bx)}
\newcommand{\CC}{C_\bx}

\newcommand{\UUU}{\mathcal{U}}
\newcommand{\VVV}{\mathcal{V}}
\newcommand{\GGG}{\mathcal{G}}
\newcommand{\HHH}{\mathcal{H}}

\newcommand{\OOO}{\mathcal{O}} 
\newcommand{\AAA}{\mathcal{A}}

\newcommand{\RRR}{\mathcal{R}} 

\newcommand{\smallr}{r}
\newcommand{\bigT}{T}
\newcommand{\bigR}{R}

\newcommand{\Vjet}{{V^{\JET}}}
\newcommand{\VVVjet}{{\VVV^{\JET}}}
\newcommand{\Ujeti}{{U_i^{\JET}}}
\newcommand{\UUUjeti}{{\UUU_i^{\JET}}}
\newcommand{\Vgmd}{{V_{\phantom{i}}^{\GMD}}}
\newcommand{\VVVgmd}{{\VVV_{\phantom{i}}^{\GMD}}}
\newcommand{\Ugmdi}{{U_i^{\GMD}}}

\newcommand{\Ugmd}{{U_{\phantom{i}}^{\GMD}}}
\newcommand{\UUUgmd}{{\UUU_{\phantom{i}}^{\GMD}}}
\newcommand{\Ugmdone}{{U_1^{\GMD}}}
\newcommand{\Ugmdtwo}{{U_2^{\GMD}}}
\newcommand{\Ugmdidagger}{{  \left( U_i^{\GMD}  \right)^\dagger    }}
\newcommand{\Ujetidagger}{{  \left( U_i^{\JET}  \right)^\dagger    }}
\newcommand{\Vgmddagger}{{  \left( V^{\GMD}  \right)^\dagger    }}
\newcommand{\Vjetdagger}{{  \left( V^{\JET}  \right)^\dagger    }}
\newcommand{\UUUgmdidagger}{{  \left( \UUU_i^{\GMD}  \right)^\dagger    }}
\newcommand{\UUUjetidagger}{{  \left( \UUU_i^{\JET}  \right)^\dagger    }}

\newcommand{\blkmat}[2]{\left\lceil   {#1}  \right\rfloor_{\otimes #2}}

\IEEEoverridecommandlockouts 

\begin{document}

\title{Modulation for MIMO Networks with Several Users}

\author{
\authorblockN{Anatoly Khina}
\authorblockA{Dept. of EE-Systems,
TAU\\
Tel Aviv, Israel \\
Email: anatolyk@eng.tau.ac.il}
\and
\authorblockN{Ayal Hitron}
\authorblockA{Dept. of EE-Systems,
TAU\\
Tel Aviv, Israel \\
Email: ayal@eng.tau.ac.il}
\and
\authorblockN{Uri Erez\authorrefmark{1}}
\authorblockA{Dept. of EE-Systems,
TAU\\
Tel Aviv, Israel \\
Email: uri@eng.tau.ac.il}
\thanks{$^*$ This work was supported in part by the U.S. - Israel Binational Science
Foundation under grant 2008/455.}
}

\maketitle

\begin{abstract}
In a recent work, a capacity-achieving scheme for the common-message two-user MIMO broadcast channel, based on single-stream coding and decoding, was described. This was obtained via a novel joint unitary triangularization which is applied to the corresponding channel matrices. In this work, the triangularization is generalized, to \emph{any} (finite) number of matrices, allowing multi-user applications. 
To that end, multiple channel uses are jointly treated, in a manner reminiscent of space-time coding.
As opposed to the two-user case, in the general case there
does not always exist a perfect (capacity-achieving) solution. However, a nearly optimal scheme
(with vanishing loss in the limit of large blocks) always exists. Common-message broadcasting is but one example of communication networks with MIMO links which can be solved using an approach coined  ``Network Modulation"; the extension beyond two links carries over to these problems.

\end{abstract}

\section{Introduction}
\label{s:intro}

Multiple-input multiple-output (MIMO) Gaussian channels are a basic building block of many communication networks, due to their potential to enhance the throughput of communication systems, and  have been extensively studied both in terms of the theoretical limits (see, e.g., \cite{Telatar99}) as well as in terms of modulation and coding schemes that allow to approach these limits. In different communication scenarios, different assumptions on the channel behavior and of the availability of channel state information  are appropriate (see \cite{gore_book,tse_book} and references therein).

A recent approach, coined ``Network Modulation''~\cite{JET:SP}, tackles the problem of conveying information over different multiple-antenna multi-terminal networks where full channel state information is available at all terminals (i.e., a fully closed-loop scenario). The approach is based on jointly triangularizing several matrices using the same unitary matrix on one side (joint encoder or decoder) and different unitary matrices on the other side (separate decoders or encoders), such that the diagonals of the resulting triangular matrices satisfy desirable properties, e.g., that they are equal. This decomposition, along with successive interference cancellation (SIC) or dirty paper coding (DPC) \cite{costa}, transforms the channels into parallel scalar additive white Gaussian noise channels (AWGN). Thus, employing this scheme along with (any) scalar codes which are good for the AWGN channel, provides ``practical'' capacity-achieving schemes, for scenarios in which the capacity is known. Furthermore, somewhat surprisingly, it has been demonstrated that the approach allows to obtain new achievable rate regions to several information-theoretic problems, such as the two-way MIMO relay problem \cite{two_way_relay} and the problem of joint source-channel coding of a source over a MIMO broadcast channel \cite{JET:SP}.

A scenario of significant importance is that of sending a common message over a MIMO Gaussian broadcast (BC) channel, henceforth the \emph{multicasting} scenario. The channel is given by
\begin{align} \label{eq:channel_model}
  \by_i = H_i \bx + \bz_i \,, \qquad i=1,2 \,,
\end{align}
where $\bx$ is the complex-valued channel input vector of length $n$ subject to a power constraint, $\by_i$ ($i=1,2$) is the output vector of user $i$ of length $m_i$, $H_i$ is the $m_i \times n$ complex channel matrix to user $i$, and $\bz_i$ is an additive circularly-symmetric complex Gaussian noise vector of length $m_i$. Without loss of generality, we assume that both the noise elements and the input signal have unit power, i.e., $z_i \sim \mathcal{CN}(0,I_{m_i})$  and $\mathbb{E} \left[ \bx^\dagger \bx \right] \leq 1$, where $\dagger$ denotes the conjugate transpose operator. It was shown in \cite{JET:SP} that it is possible to jointly triangularize the channel matrices $H_1$ and $H_2$ using unitary matrices, such that the ratio between the resulting diagonals is constant. This in turn allows to achieve the common-message capacity using single-stream encoding and decoding of standard AWGN codes along with SIC (much like in V-BLAST transmission for a single user \cite{Wolniansky_V-BLAST}). As we recall in the sequel, the problem of  multicasting over several MIMO channels is tightly connected to the problems of universal coding over parallel channels, as well as rateless coding for Gaussian channels. Thus, the results we derive are relevant also to the latter problems.  

The joint triangularization of~\cite{JET:SP} was limited to only two matrices, and hence, only two-user multicasting (or ``perfect two-rate'' in the rateless problem \cite{relay_mimo_itw}) could be treated. The aim of the current work is to generalize the network modulation approach to more than two users. This is done by utilizing multiple uses of the channel, reminiscent of space-time coding techniques~\cite{alamouti,stc}.

\section{Background: Network Modulation}
In this section we recall the joint triangularization of two matrices \cite{JET:SP}, and its application to the two-user multicasting problem. We then demonstrate the relevance of the scheme to the special case of a two-rate scalar Gaussian rateless problem.

\subsection{Unitary Matrix Triangularization}

The network modulation approach is based on several forms of matrix decompositions, one of which is the \emph{geometric mean decomposition} (GMD) \cite{GMD}.  For simplicity,  we will only consider the decomposition of \emph{square} matrices throughout this work. As we show in the sequel, this does not pose any restriction on the communication problems addressed. The GMD  \cite{GMD} of an $n \times n$ matrix $A$ is given by:
    \begin{align} \label{eq:gmd}
        A &= U \bigT V^\dagger \,,
    \end{align}
where $U$ and $V$ are $n \times n$ unitary matrices, and $\bigT$ is an upper-triangular $n \times n$ matrix such that all its diagonal values equal $\lambda$, where $\lambda$ is a real-valued non-negative number.

Building on the GMD, the following decomposition, which will be referred to as Joint Equi-diagonal Triangularization (JET),  was introduced in \cite{JET:SP}. Let  $A_1$ and $A_2$ be two complex matrices of dimensions $n \times n$ such that \mbox{$\det (A) = \det (B)$}. Then, the joint triangularization of $A_1$ and $A_2$ is given by:
    \begin{align}
	\nonumber        A_1 &= U_1 \bigR_1 V^\dagger \\
	\label{eq:jet2}        A_2 &= U_2 \bigR_2 V^\dagger \,,
    \end{align}
where $U_1,U_2,V$ are $n \times n$ unitary matrices,     and $\bigR_1,\bigR_2$ are upper-triangular
 $n \times n$ matrices with \emph{the same} real-valued, non-negative diagonal values, namely,
$$
    \left[ \bigR_1 \right]_{ii} = \left[ \bigR_2 \right]_{ii} \,\quad \forall i=1,\ldots,n \,.
$$

\subsection{MIMO Multicast Scheme}
\label{ss:old_scheme}
\label{ss:rateless}

We now recall how the JET decomposition can be used to obtain a practical scheme for transmitting a common message over a MIMO Gaussian BC  with two receivers, as described by \eqref{eq:channel_model}. Define the mutual information between a Gaussian input vector $\bx$, having a covariance matrix  $\CC \triangleq  \mathbb{E} \left[ \bx \bx^\dagger \right]$, and the channel output $\by_i$, by 
\begin{equation} \label{eq:information}
\Capacity{i} \triangleq \log \det \left( I + H_i \CC H_i^\dagger \right) \,.
\end{equation}
The \emph{common-message capacity} is given by the (worst-case) compound-channel capacity expression (see, e.g.,~\cite{BlackwellBreimanThomasian59}):
\begin{align}
      \label{eq:Compound_Capacity}
      C = \max_{\CC :\, \mathrm{tr} (\CC) \leq 1} \min_{i=1,2} \Capacity{i} \,.
\end{align}

Let $\CC$ be an admissible covariance matrix, and assume for simplicity that $\Capacity{1}=\Capacity{2}=R$. The following scheme~\cite{JET:SP} achieves the rate $R$.

Define the following \emph{augmented matrices}:
\begin{equation*} 
	\tilde{G}_i \triangleq 
	\left[
	      \begin{array}{c} F_i \\ I_n \end{array}
	\right]\,,
\end{equation*}
where $ F_i \triangleq  H_i \sqrt{\CC} $ and $I_n$ is the $n \times n$ identity matrix.

Next, the matrices $\tilde{G}_i$ are transformed into square matrices, by means of the QR decomposition:
\begin{equation} \label{eq:G_matrix}
	\tilde{G}_i = Q_i G_i\,,
\end{equation}
where $Q_i$ is an $(m_i+n) \times n$ matrix with orthonormal columns and $G_i$ is an $n \times n$ upper-triangular matrix with real-valued positive diagonal elements. Now,  assuming that $\Capacity{1}=\Capacity{2}=R$,
this implies \cite[Proposition~1]{JET:SP}: 
$$
    \det( G_1 )= \det ( G_2 ) = 2^\frac{R}{2} \,.
$$
Therefore, $G_1,G_2$ can be jointly triangularized using the JET:
\begin{equation} \label{eq:jet_of_old_scheme}
      G_i =  U_i \bigR_i V^\dagger \,, \qquad i=1,2\,, 
\end{equation}
where $\bigR_1$ and $\bigR_2$ are upper-triangular, having the same diagonal elements. The  transmission scheme is as follows:
\begin{enumerate}
 \item
Construct $n$ optimal codes for scalar AWGN channels. The \blath{$k$} codebook is designed for a SISO AWGN channel with a rate $2 \log \smallr_k$, where $\smallr_k$ is the \blath{$k$} diagonal element of $\bigR_1$ (and also of $\bigR_2$).

\item
In each channel use, an $n$-length vector $\tilde\bx$ is formed using one sample from each codebook. The transmitted vector $\bx$ is then obtained using the following precoder:
\begin{equation} \label{eq:bx_tilde_to_bx}
	\bx = \sqrt{\CC} V \tilde\bx \,.
\end{equation}

\item At the receiving ends, the \blath{$i$} user calculates
\begin{equation} \label{eq:scheme_receiver}
	\tilde\by_i = U_i^\dagger \tilde{Q}_i^\dagger \by_i\,,
\end{equation}
where $\tilde{Q}_i$ consists of the first $n$ rows of $Q_i$.

\item Finally, the codebooks are decoded using SIC, starting from the \blath{$n$}  codeword and ending with the first one: The \blath{$n$} codeword is decoded first, using the \blath{$n$} element of $\tilde\by_i$,
treating the other codewords as AWGN. The  effect of the \blath{$n$} element of $\tilde{\bx}$ is then subtracted out from the remaining elements of $\tilde{\by}$. Next, the \blath{$(n-1)$} codeword is decoded, using the \blath{$(n-1)$} element of $\tilde\by_i$ - and so forth.
\end{enumerate}
The optimality of this scheme was proved in~\cite[Sec.~IV]{JET:SP}.

\begin{example}[Application to the two-rate rateless problem]
Consider the scalar Gaussian rateless problem defined in \cite{etw}:
$$
      y_m = \alpha x_m + z_m\,, \qquad m = 1,2,\ldots,M \,.
$$
The gain $\alpha$ is known only to the receiver, and can take one of $M$ possible values, such that  a gain of $\alpha_m$ implies that the 
message should be decodable using $m$ received blocks: \footnote{Alternatively, this can be viewed as a scheme that works for every value of $\alpha$, but designed to be optimal only for $M$ specific values.}
$$
	R = m \log (1 + | \alpha_m |^2 ) \,, \qquad m=1,2,\ldots,M \,.
$$
Specializing the problem to the case of one (possible) incremental redundancy block ($M=2$), the perfect two-rate rateless problem can be viewed as a 2-user MIMO-BC channel with channel matrices
        \begin{align*}
            H_1 = \left(  \begin{array}{cc} \alpha_1 & 0  \\ 0  & 0 \end{array}\right)
        	,\quad
            H_2 = \left(  \begin{array}{cc} \alpha_2 & 0  \\ 0  & \alpha_2 \end{array}\right) \,.
        \end{align*}	
Applying the scheme of \secref{ss:old_scheme} yields the following precoding matrix \cite{relay_mimo_itw}:
        \begin{align*}
            V =
            \sqrt{\frac{1}{2^{R/2}+1}}
            \left(
              \begin{array}{cc}
                1 & 2^{R/4} \\
                2^{R/4} & -1 \\
              \end{array}
            \right)  \,,
        \end{align*}
which coincides with the result in~\cite[Section~III]{etw}.

Erez, Trott and Wornell \cite{etw} also treated the case of $M=L=3$, and found a condition for which a ``perfect'' scheme exists. 
In the sequel we will shed light on this condition. 

\end{example}

\section{Joint Triangularization of Many Matrices}
\label{ss:joint_decomposition}

In this section we extend the network modulation technique to a any finite number of users,
using a \emph{recursion principle}.
Specifically, given $K$ matrices $G_1,\ldots,G_{K}$, we wish to find $K$ matrices with orthonormal columns
$U_1,\ldots,U_k$, and another such matrix $V$, such that the matrices $\bigR_i \triangleq U_i^\dagger G_i V$
are upper-triangular, having \emph{equal} diagonals.
We shall refer to this as $K$-matrix JET, or simply $K$-JET.

The proof of the existence of a JET decomposition of two matrices $G_1$ and $G_2$ \cite{JET:SP} is based 
upon applying the GMD \eqref{eq:gmd} to the single matrix  $G_1 G_2^{-1}$.
Similarly, we show in the following lemma that  $(K+1)$-JET is equivalent to simultaneous GMD of $K$ matrices, which will be referred to as $K$-GMD.
\begin{lemma}  \label{lem:k_to_k_plus_one}
Let $G_1,\ldots,G_{K+1}$ be $n \times n$ complex valued matrices with equal determinants, and
define the $K$ matrices:
\begin{align} \label{eq:definition_of_a}
			A_i = G_i G_{K+1}^{-1}  \,, \qquad i=1,\ldots,K \,.
\end{align}
Then, there exist $K+1$
matrices with orthonormal columns $U_1,\ldots,U_{K+1}$, of dimensions $n \times m$, such that
\begin{align} \label{eq:jet_of_a}
	U_i^\dagger A_i  U_{K+1} = \bigT_i \,,\qquad  i=1,\ldots,K \,,
\end{align}
where $\left\{ \bigT_i \right\} $ are upper-triangular with all diagonal entries equal to $1$,
\emph{if and only if} there exists an  $n \times m$ matrix $V$ with orthonormal columns, such that
$$
	U_i^\dagger G_i V = \bigR_i \,,\qquad i=1,\ldots,K+1 \,,
$$
where $\left\{ \bigR_i \right\}$ are upper-triangular with equal diagonals.
\end{lemma}

\begin{IEEEproof}
See a constructive proof in \appref{app:proof_k_to_k_plus_one}.
\end{IEEEproof}

\begin{remark}
    Constructing matrices with constant diagonals could be advantageous in practice, as this corresponds to equal gains of all the resulting sub-channels, and hence enables to use the \emph{same} (single) codebook over all of them.
\end{remark}

\vspace{.3\baselineskip}
We are thus left with the task of  performing $K$-GMD to $K$ matrices. In \secref{s:exact_kgmd} we state sufficient and necessary conditions for the existence of the above decomposition for the special case of two real-valued  $2 \times 2$ matrices. We will then, in \secref{s:space_time_triangularization}, present a different approach, involving joint triangularization of block-diagonal matrices, which enables a nearly-optimal network-modulation scheme, even when exact triangularization is not possible.

\section{Exact Triangularization with constant diagonals of two real-valued $2 \times 2$ matrices}
\label{s:exact_kgmd}

We now provide a  necessary  and sufficient condition for the existence of $2$-GMD for \emph{real-valued} $2 \times 2$ matrices. 

\begin{thm}[2-GMD for $2 \times 2$ real-valued matrices] \label{thm:perfect_two_on_two}
  Let $A_1$ and $A_2$ be \emph{real-valued} $2 \times 2$ matrices with determinants equal to $1$.   Apply (any) JET decomposition to them: \footnote{The JET decomposition is, in general, not unique.}
\begin{equation} \label{eq:jet_of_augmented}
      A_i =  \Ujeti \bigR_i \Vjetdagger \,, \qquad i=1,2\,,
\end{equation}
    where:
$$
      \bigR_i = \left( \begin{array}{cc}
		      \smallr_1 & x_i \\ 0 & \smallr_2
                   \end{array} \right) \,.
$$
Then, there exist three complex-valued $2 \times 2$ unitary matrices $\Ugmdone,\Ugmdtwo,\Vgmd$ such that:
$$
      \Ugmdidagger A_i \Vgmd =  \left( \begin{array}{cc}
		      1 & * \\ 0 & 1
                   \end{array} \right)
$$
\emph{if and only if} the following inequality is satisfied:
\begin{equation} \label{eq:condition}
      \smallr_2 \left( \frac{x_1 + x_2}{2} \right)^2 \leq \smallr_2 + \frac{x_1 x_2}{\smallr_1 - \smallr_2}.
\end{equation}
Without loss of generality, we can assume that the solution is of the form:
\begin{equation} \label{eq:complex_orthogonal_design}
      \Vgmd = \left( \begin{array}{cc}
	    s_1 & \phantom{-}s_2  \\
	    s_2^* & -s_1^* \\		
                   \end{array} \right)\,.
\end{equation}
\end{thm}
\begin{IEEEproof}
The proof is straightforward, and is given in \appref{app:real_proof}.
\end{IEEEproof}

\begin{remark}
Although this theorem is valid only for \emph{real-valued} matrices $A_1$ and $A_2$, the resulting unitary matrices $U_1,U_2,$ and $V$ are, in general, \emph{complex-valued}. In  \secref{ss:restatement} we bring a restatement of the theorem, which involves only real-valued orthogonal transformations.
\end{remark}

\begin{remark}
This theorem can be applied to the three-rate rateless problem defined in \secref{ss:rateless}. This yields a condition for the existence of a perfect scheme, namely, $  R \leq 6 \log \left( \frac{3 + \sqrt{5}}{2}      \right) \approx 8.331$, as in  \cite{etw}. The details are given  in \appref{app:rateless}.
\end{remark}

\section{Space-Time Triangularization}
\label{s:space_time_triangularization}

As indicated by \thrmref{thm:perfect_two_on_two}, joint triangularization with constant diagonal values is not always possible. However, even when the condition for joint triangularization does not hold, we can still perform nearly-optimal network modulation, by utilizing multiple uses of the same channel realization. The idea of mixing the same symbols between multiple channel uses has much in common with Space-Time Codes \cite{alamouti,stc}.

\subsection{Restatement of \thrmref{thm:perfect_two_on_two}}
\label{ss:restatement}

In order to introduce the space-time like structure, we start by a restatement of \thrmref{thm:perfect_two_on_two}.

Recall the two-user common-message broadcast MIMO channel \eqref{eq:channel_model} with two transmit antennas ($n=2$), and a general number of antennas $m_i$ at each receiver. We now utilize transmission in two consecutive time instances (as in \cite{alamouti}). This is equivalent to sending extended symbols over the following \emph{extended channel}:
$$
	\bY_i = \HHH_i \bX + \bZ_i \,, \qquad i=1,2  \,.
$$
The extended vectors $\bX,\bY,\bZ$ are composed of two ``physical'' input, output, and noise vectors, respectively, and $\HHH_i$ is the $(2m_i) \times 4$ \emph{extended channel matrix} defined as ($i=1,2$)
\begin{equation} \label{eq:extended_two_H}
 \HHH_i = \blkmat{H_i}{2} \,
\end{equation}
where $ \blkmat{A}{N} $ denotes the Kronecker product $I_N \otimes A$, viz.\ a block-diagonal matrix with $N$ blocks of $A$ on its diagonal:
$$
    \blkmat{A}{N} \triangleq \left( \begin{array}{cccc} A & 0  & \cdots & 0 \\ 0 & A & \cdots & 0 \\ \vdots & \vdots & \ddots & \vdots \\ 0 & 0 & \cdots & A
                   \end{array} \right)  \,.
$$
The power constraint now becomes $\mathbb{E} \left[ \bX^T \bX \right] \leq 2$.

Let $\CC$ be a covariance matrix satisfying $\trace \CC \leq 1$, and define the augmented matrices $G_i$ as in
\eqref{eq:G_matrix}. Following \lemref{lem:k_to_k_plus_one}, we define the two $2 \times 2$ matrices:
$$
    A_1 \triangleq G_1 G_3^{-1} \,, \; A_2 \triangleq G_2 G_3^{-1} \,.
$$
Also define the following $4 \times 4$ extended matrices ($i=1,2$):
\begin{equation} \label{eq:extended_two}
    \GGG_i \triangleq  \blkmat{G_i}{2} \,, \;  \AAA_i \triangleq  \blkmat{A_i}{2} \,.
\end{equation}

Since the matrices $A_1$ and $A_2 $ are real-valued matrices, we can obtain $2$-GMD of the matrices $\AAA_1$ and $\AAA_2$ under the same conditions as in \thrmref{thm:perfect_two_on_two}, such that all the involved unitary transformations become \emph{real-valued}. Following \lemref{lem:k_to_k_plus_one}, this yields a 3-JET of the three matrices $\GGG_1,\GGG_2,\GGG_3$:
$$
      \GGG_i = \UUU_i \RRR_i \VVV^\dagger \,,
$$
where $\RRR_i$ are upper triangular with equals diagonals.

In particular, the complex precoding matrix $V^{\mathrm{GMD}}$ given by \eqref{eq:complex_orthogonal_design} 
implies the following (real) orthogonal space-time block code structure of $\VVV^{\mathrm{GMD}}$ \cite{orthogonal_design}:
$$
      \VVV^{\mathrm{GMD}} = \left( \begin{array}{cccc}
		      \phantom{-}a & -c & \phantom{-}b & \phantom{-}d \\
		      \phantom{-}b & \phantom{-}d & -a & \phantom{-}c \\
			\phantom{-}c & \phantom{-}a & \phantom{-}d & -b \\
			  -d & \phantom{-}b & \phantom{-}c & \phantom{-}a
                   \end{array} \right) \,.
$$

The same scheme as in \secref{ss:old_scheme} can now be employed, such that the two channel uses are effectively transformed into four scalar AWGN channels, having the same capacities for all three users. Note that the matrix $\tilde{Q}_i$ of \eqref{eq:scheme_receiver} is replaced with its extended version, $ \blkmat{ \tilde{Q}_i }{2} $.

\subsection{Nearly-Optimal $2$-GMD}
\label{s:asymptotic_triangularization}

We now show how to utilize a space-time structure in order to obtain nearly-optimal joint triangularization of two matrices, such that the resulting triangular matrices have a constant diagonal. This method will later be generalized to any number of matrices, using \lemref{lem:k_to_k_plus_one}. The resulting scheme becomes asymptotically optimal for large values of $N$, where $N$ is the number of channel uses assembled together for the purpose of joint decomposition. Note that the proposed scheme is nearly optimal for \emph{any} two complex-valued channels $H_i$ (and not restricted to real-valued matrices, in contrast to the perfect construction of \thrmref{thm:perfect_two_on_two}).

\begin{thm}[Nearly-Optimal $2$-GMD]   \label{thm:n_n_asymptotical}
Let $A_1$ and $A_2$ be two complex-valued $n \times n$ matrices, and define the following $nN \times nN$ extended matrices:
\begin{equation} \label{eq:extended_n}
  \AAA_i = \blkmat{A_i}{N} \,, \quad i=1,2 ,\,.
\end{equation}
Then there exist three $nN \times n \big(N-(n-1) \big)$ matrices $\UUU_1,\UUU_2,\VVV$ with orthonormal columns, such that:
$$
      \UUU_i^\dagger \AAA_i \VVV = \left( \begin{array}{ccccc}
		      1 & * & \cdots & * & * \\
		      0 & 1 & \cdots & * & *\\
		      \vdots & \vdots & \ddots & \vdots & \vdots \\
		      0 & 0 & \cdots & 1 & * \\
		      0 & 0 & \cdots & 0 & 1
                   \end{array} \right) \,, \qquad i=1,2 \,.
$$
\end{thm}

By using this decomposition together with \lemref{lem:k_to_k_plus_one}, the same scheme as in \secref{ss:old_scheme} can be employed, such that the $N$ channel uses are effectively transformed into \mbox{$n(N-n+1)$} scalar AWGN channels. The sum of the capacities of these channels tends to the capacity of the original channel for large values of $N$, where the only loss~comes from edge effects (truncating the extreme $n(n-1)$ elements).

The full proof of the theorem is given in \appref{app:large_n_proof}. The main idea of the proof is demonstrated by the proof for the $2 \times 2$ case, presented next.

\begin{IEEEproof}[Proof of \thrmref{thm:n_n_asymptotical} for $n=2$]
We start by jointly triangularizing the matrices $A_1$ and $A_2$:
\begin{eqnarray} \label{eq:stage_a}
	\Ujetidagger A_i \Vjet & = & \left( \begin{array}{cc}
	                     \smallr_1 & x_i \\ 0 & \smallr_2
	                 \end{array} \right)
\end{eqnarray}
where $\smallr_1 \smallr_2 = 1$. We now apply the decomposition \eqref{eq:stage_a} to each block separately, using:
$$
  \UUUjeti = \blkmat{ \Ujeti }{N}   \,,\; \VVVjet = \blkmat{ \Vjet }{N} \,,
$$
which yields the matrices
\begin{equation} \label{eq:big_btb} 
      \UUUjetidagger \AAA_i \VVVjet = \left(
\begin{array}{ccccccc} \cline{1-2}
	 \multicolumn{1}{|c}{\smallr_1} & \multicolumn{1}{c|}{x_i} & 0 & 0 & \cdots & 0 & 0 \\ \cdashline{2-3}
	 \multicolumn{1}{|c}{0}    &  \multicolumn{1}{:c|}{\smallr_2} & 0 & \multicolumn{1}{:c}{0} & \cdots & 0 & 0 \\ \cline{1-4}
	  0 & \multicolumn{1}{:c}{0} & \multicolumn{1}{|c}{\smallr_1} & \multicolumn{1}{:c|}{x_i}  & \cdots& 0 & 0  \\ \cdashline{2-3}
	  0 & 0 &  \multicolumn{1}{|c}{0}    & \multicolumn{1}{c|}{\smallr_2} &\cdots &  0 & 0  \\ \cline{3-4}
	  \vdots  &  \vdots & \vdots  &  \vdots   & \ddots &  \vdots  &  \vdots   \\	    \cline{6-7}
	  0 & 0 & 0 & 0 &\cdots & \multicolumn{1}{|c}{\smallr_1} & \multicolumn{1}{c|}{x_i} \\
	  0 & 0 & 0 & 0 &\cdots &  \multicolumn{1}{|c}{0}    & \multicolumn{1}{c|}{\smallr_2} \\ \cline{6-7}
\end{array}
\right)\,.
\end{equation}
Note that the sub-matrix
$$
      \Lambda = \left(
      \begin{array}{:cc:} \hdashline
	\smallr_2 & 0 \\
	  0 & \smallr_1 \\ \hdashline
      \end{array} \right)
$$
does not depend on $i$, and therefore it can be decomposed using the GMD \eqref{eq:gmd}, $      \Lambda = \Ugmd \bigT \Vgmddagger $, where $\bigT$ is upper-triangular with only $1$s on the diagonal. We use this decomposition to construct a second transformation -- only this time it is not be applied on each block separately, but rather ``mixes'' pairs of consecutive blocks, using:
$$
  \UUUgmd = 
    \left(
\begin{array}{cccc}
0 & 0 & \cdots & 0 \\
\multicolumn{4}{c}{ \multirow{2}{*}{$\blkmat{\Ugmd}{(N-1)}$} } \\
\multicolumn{4}{c}{ } \\
0 & 0 & \cdots & 0 \end{array} \right),  
\VVVgmd = 
    \left(
\begin{array}{cccc}
0 & 0 & \cdots & 0 \\
\multicolumn{4}{c}{ \multirow{2}{*}{$\blkmat{\Vgmd}{(N-1)}$} } \\
\multicolumn{4}{c}{ } \\ 0 & 0 & \cdots & 0 \end{array} \right)\,.
$$

Applying this transformation to \eqref{eq:big_btb} yields the following $(2N-2) \times (2N-2)$ upper-triangular matrix:
\begin{equation*}
      \UUU_i^\dagger \AAA_i \VVV =  \left(
\begin{array}{ccccc}	   \cdashline{1-2}	
	   \multicolumn{1}{:c}{1}  & \multicolumn{1}{c:}{*} &  \cdots & * & * \\
 	   \multicolumn{1}{:c}{0}  & \multicolumn{1}{c:}{1} &  \cdots& * & *  \\  \cdashline{1-2}	
	     \vdots   & \vdots &  \ddots  &  \vdots & \vdots  \\ \cdashline{4-5}
	    0 &0 & \cdots & \multicolumn{1}{:c}{1} & \multicolumn{1}{c:}{*} \\
	    0 &0 & \cdots &  \multicolumn{1}{:c}{0} & \multicolumn{1}{c:}{1} \\ \cdashline{4-5}
\end{array}
\right),
\end{equation*}
where
$ \UUU_i \triangleq \UUUjeti \UUUgmd $
and
$ \VVV \triangleq \VVVjet \VVVgmd $.
\end{IEEEproof}

\subsection{Nearly-Optimal $K$-GMD}

By using \lemref{lem:k_to_k_plus_one}, we can generalize \thrmref{thm:n_n_asymptotical} to any number of users, as follows:

\begin{thm}[Nearly-Optimal $K$-GMD]   \label{thm:n_n_asymptotical_k_users}
Let $A_1, \ldots , A_K$ be $K$ complex-valued $n \times n$ matrices with determinants equal to $1$, and define $\AAA_1, \ldots, \AAA_K$ as in \eqref{eq:extended_n}. Then there exist $K+1$ matrices $\UUU_1,\ldots,\UUU_K,\VVV$, with orthonormal columns, such that:
$$
      \UUU_i^\dagger \AAA_i \VVV =  \left( \begin{array}{ccccc}
		      1 & * & \cdots & * & * \\
		      0 & 1 & \cdots & * & *\\
		      \vdots & \vdots & \ddots & \vdots & \vdots \\
		      0 & 0 & \cdots & 1 & * \\
		      0 & 0 & \cdots & 0 & 1
                   \end{array} \right) \,, \qquad  i=1,\ldots,K \,.
$$
\end{thm}
\begin{IEEEproof}
A sketch of the proof is given in Appendix~\ref{app:general_case}.
\end{IEEEproof}

\section{Discussion}
\label{s:conc}
\thrmref{thm:perfect_two_on_two} provides sufficient and necessary conditions for joint GMD of two \emph{real-valued} $2 \times 2$ matrices. This naturally raises the question of how this condition can be  carried over to the complex-valued case, and to general dimensions $n \times n$.

Furthermore, we demonstrated that (exact) $K$-GMD, not using any space-time structure, is not always possible. Nevertheless, \emph{nearly}-optimal communication schemes can always be constructed, which become optimal in the limit of large $N$. It remains an open question whether an exact triangularization can be obtained using only a finite number of channel uses.

\section*{Acknowledgements}
The authors would like to thank Yuval Kochman for constant help throughout this work.

\appendices

\section{Proof of  \lemref{lem:k_to_k_plus_one}}
\label{app:proof_k_to_k_plus_one}

\begin{IEEEproof}[Proof of \lemref{lem:k_to_k_plus_one}]
The direct part  holds trivially. We are therefore left with the task of proving the converse part. We start with the QR decomposition
$	G_{K+1}^{-1} U_{K+1} = V S $,
where $V$ is of dimensions $n \times m$ with orthonormal columns, and $S$ is an \mbox{$m \times m$}  upper-triangular matrix. Thus, using \eqref{eq:definition_of_a} and \eqref{eq:jet_of_a}, we obtain the following equalities:
\begin{align*}
&   U_i^\dagger G_i V S            = \bigT_i  \,, \qquad i=1,\ldots,K \\
&   U_{K+1}^\dagger G_{K+1} V S    = I \,.
\end{align*}
Multiplying by $S^{-1}$ on the right yields:
\begin{align*}
&    U_i^\dagger G_i V         = \bigT_i S^{-1}  \,, \qquad i=1,\ldots,K \\
&    U_{K+1}^\dagger G_{K+1} V = S^{-1} \,.
\end{align*}
Since $T_i$ are upper-triangular with only $1$s on the diagonal, the matrices
$ \bigR_i  \triangleq \bigT_i S^{-1} $  ($i=1,\ldots,K$) and
$ \bigR_{K+1} \triangleq S^{-1} $
have equal diagonals, which completes the proof.
\end{IEEEproof}

\section{Sketch of Proof of Theorem~\ref{thm:n_n_asymptotical_k_users}}
\label{app:general_case}
\emph{Proof Idea:}
The theorem has already been proved for the special case of $K=2$. For larger values of $K$ we prove by induction, applying repeatedly \lemref{lem:k_to_k_plus_one} and of \thrmref{thm:n_n_asymptotical}:
\begin{enumerate}
	\item According to \lemref{lem:k_to_k_plus_one}, performing $K$-GMD is equivalent to $(K+1)$-JET. We can thus
		transform $K$ upper-triangular matrices with \emph{constant} diagonal values
		into $K+1$  upper-triangular matrices of the same size, $\bigR_1, \ldots, \bigR_{K+1}$ with \emph{equal} diagonals.
	
	\item Given the matrices $\bigR_1, \ldots, \bigR_{K+1}$, construct the block-diagonal \emph{extended} matrices $\RRR_i$,
as in \eqref{eq:extended_n}. Using the technique of \thrmref{thm:n_n_asymptotical}, we construct matrices with orthonormal columns,
$\UUU^{(K+1)}_1, \ldots , \UUU^{(K+1)}_{K+1},\VVV^{(K+1)}$, such that  the matrices
$
		\left({\UUU^{(K+1)}_i}\right)^\dagger \RRR_i \VVV^{(K+1)}
$
are upper-triangular, with constant diagonals. Finally, the loss in rate could be made arbitrarily small by taking $N$ to be sufficiently large.

\end{enumerate}

\section{Condition for $2$-GMD of Real-Valued $2 \times 2$ Matrices}
\label{app:real_proof}

We now prove the necessary and sufficient condition for the existence of joint-triangularization of two $2 \times 2$ real-valued matrices.

\begin{IEEEproof}[Proof of \thrmref{thm:perfect_two_on_two}]
Let $A_1$ and $A_2$ be real-valued $2 \times 2$ matrices with determinants equal to $1$.
Apply the JET decomposition to these matrices, to obtain
\begin{equation} \label{eq:blablabla}
      A_i =  \Ujeti \bigR_i \Vjetdagger \,, \qquad i=1,2,
\end{equation}
    where:
$$
      \bigR_i = \left( \begin{array}{cc}
		      \smallr_1 & x_i \\ 0 & \smallr_2
                   \end{array} \right)
$$
such that $\smallr_1 \smallr_2 = 1$. The matrices $\Ujeti,\Vjet$ are real-valued unitary matrices, and we can assume without loss of generality that $\det \left( \Vjet \right) = 1$.

If there exist three complex-valued $2 \times 2$ unitary matrices ${U}_1,{U}_2,{V}$ such that:
\begin{equation} \label{eq:decomposition_again}
      {U}_i^\dagger A_i {V} = \left( \begin{array}{cc}
		      1 & * \\ 0 & 1
                   \end{array} \right),
\end{equation}
then according to \eqref{eq:blablabla},
$$
      \Ugmdidagger \bigR_i \Vgmd = \left( \begin{array}{cc}
		      1 & * \\ 0 & 1
                   \end{array} \right)\,,
$$
where
\begin{eqnarray*}
	\Ugmdi & = & \Ujetidagger U_i \\
	\Vgmd & = & \Vjetdagger V \,.
\end{eqnarray*}
Denote the entries of the first column of $\Vgmd$ by $s_1$ and $s_2^*$, i.e., 
$$
      \Vgmd = \left(
	  \begin{array}{cc}
	    s_1 & * \\ s_2^* & *
	  \end{array}
\right)\,,
$$
where $| s_1|^2 + |s_2|^2 = 1$. The first column of $\bigR_1 \Vgmd$ and of $\bigR_2 \Vgmd$ is therefore:
$$
      \bigR_i \Vgmd = \left(
	  \begin{array}{cc}
	    \smallr_1 s_1 + x_i s_2^* & * \\ \smallr_2 s_2^* & *
	  \end{array}
\right) \,,\qquad i=1,2\,,
$$
where $x_1$ and $x_2$ denote the off-diagonal elements of $R_1$ and $R_2$ respectively.
These two columns must have a norm of $1$, namely:
$$
      \left|     \smallr_1s_1 + x_is_2^*      \right|^2 + \left|     \smallr_2s_2^*      \right|^2 = 1\,,\qquad i=1,2\,.
$$
Since $\smallr_1,\smallr_2,x$ are real-valued, $s_1$ and $s_2$ must satisfy the following three equations:
\begin{eqnarray*}
    |s_1|^2 + |s_2|^2 & = & 1 \\
 \smallr_1^2 |s_1|^2 + (x_1^2+\smallr_2^2) |s_2|^2 + 2\smallr_1x_1 \mathrm{Re}(s_1s_2) & = & 1 \\
 \smallr_1^2 |s_1|^2 + (x_2^2+\smallr_2^2) |s_2|^2 + 2\smallr_1x_2 \mathrm{Re}(s_1s_2) & = & 1 \,.
\end{eqnarray*}
Denoting $
      \alpha \equiv \frac{s_1}{s_2^*}
$, and substituting $\alpha$ in these equations, results in:
\begin{eqnarray}
\label{eq:abc1}    1 + |\alpha|^2 & = & \frac{1}{|s_2|^2} \\
\label{eq:abc2}    \smallr_1^2 |\alpha|^2 + (x_1^2+r_2^2) + 2\smallr_1x_1 \mathrm{Re}(\alpha) & = & \frac{1}{|s_2|^2} \\
\label{eq:abc3}    \smallr_1^2 |\alpha|^2 + (x_2^2+r_2^2) + 2\smallr_1x_2 \mathrm{Re}(\alpha) & = & \frac{1}{|s_2|^2} \,.
\end{eqnarray}
Subtracting \eqref{eq:abc3} from \eqref{eq:abc2} yields:
$$
      (x_1+x_2) + 2 \smallr_1 \mathrm{Re}(\alpha) = 0\,,
$$
So we have:
\begin{eqnarray}
\label{eq:ttt1}         |\alpha|^2 & = & \frac{1}{|s_2|^2} - 1\\
\label{eq:ttt2}   \mathrm{Re}(\alpha) & = & - \left( \frac{x_1 + x_2}{2} \right) \smallr_2 \\
\label{eq:ttt3}    \left( \mathrm{Re}(\alpha) \right) ^2 & = & \left( \frac{x_1 + x_2}{2} \right)^2 \smallr_2^2 \\
\label{eq:ttt4}       \left( \mathrm{Im}(\alpha) \right)^2 & = &\frac{1}{|s_2|^2} - 1 - \left( \frac{x_1 + x_2}{2} \right)^2 \smallr_2^2 \,.
\end{eqnarray}
Thus, equation \eqref{eq:abc2} and \eqref{eq:abc3} become:
\begin{eqnarray*}
|s_2|^2 & = & \frac{\smallr_1^2 - 1}{\smallr_1^2 - \smallr_2^2 + x_1 x_2} \,,
\end{eqnarray*}
and therefore equation  \eqref{eq:ttt3}  becomes
\begin{eqnarray*}
 \left( \mathrm{Im}(\alpha) \right)^2 & = & \frac{\smallr_1^2 -\smallr_2^2 + x_1 x_2}{\smallr_1^2 - 1} - 1 - \left( \frac{x_1 + x_2}{2} \right)^2 \smallr_2^2 \,.
\end{eqnarray*}
Thus, the following conditions are necessary and sufficient for the existence of a solution:
\begin{eqnarray*}
  \frac{\smallr_1^2 - 1}{\smallr_1^2 -\smallr_2^2 + x_1 x_2}  & \geq & 0 \\
 \frac{\smallr_1^2 -\smallr_2^2 + x_1 x_2}{\smallr_1^2 - 1} - 1 - \left( \frac{x_1 + x_2}{2} \right)^2 \smallr_2^2 & \geq & 0     \,,
\end{eqnarray*}
which are equivalent to
\begin{equation} \label{eq:condition_again}
    \smallr_2 \left( \frac{x_1 + x_2}{2} \right)^2 \leq \smallr_2 + \frac{x_1 x_2}{\smallr_1 - \smallr_2} \,.
\end{equation}

This proves that \eqref{eq:condition_again} is a  \emph{necessary} condition for the existence of the decomposition   \eqref{eq:decomposition_again}.

Now, assume that this condition holds, and define the matrix:
$$
      \Vgmd = \left(
	  \begin{array}{cc}
	    s_1 & s_2 \\ s_2^* & -s_1^*
	  \end{array}
\right)\,.
$$
We now apply the QR decomposition to the matrices  $\bigR_1 \Vgmd$ and $\bigR_2 \Vgmd$:
\begin{eqnarray} \label{eq:bla_with_a_b_c}
  \Ugmdidagger \bigR_i \Vgmd & = & \left( \begin{array}{cc}
							  a_i & b_i \\ 0 & c_i
						 \end{array}
			  \right) \,.
\end{eqnarray}
The first columns of both $\bigR_1 \Vgmd$ and $\bigR_2 \Vgmd$ have norms equal to $1$.
Therefore, from the construction of the QR decomposition, it follows that
$a_1=a_2=1$. Consequently, since both matrices have a unit determinant,
$c_1=c_2=1$ must hold as well.
Thus, \eqref{eq:bla_with_a_b_c} becomes:
\begin{eqnarray*}
  \Ugmdidagger \bigR_i \Vgmd & = & \left( \begin{array}{cc}
							  1 & * \\ 0 & 1
						 \end{array}
			  \right) \,,
\end{eqnarray*}
and therefore,
\begin{eqnarray*}
  A_i & = &  U_i \left( \begin{array}{cc}
							  1 & * \\ 0 & 1
						 \end{array}
			  \right)  V^\dagger    \,,
\end{eqnarray*}
where
\begin{eqnarray*}
      U_i & = & \Ujeti \Ugmdi \\
	V & = &  \Vjet \Vgmd  \,.
\end{eqnarray*}
Furthermore, since the matrix $\Vgmd$ is of the form
$$
      \Vgmd = \left(
	  \begin{array}{cc}
	    s_1 & s_2 \\ s_2^* & -s_1^*
	  \end{array}
\right)\,,
$$
and $\Vjet$ is a real-valued unitary matrix with unit determinant, it is easy to see that the matrix $V$ is also of the form
$$
      V = \left(
	  \begin{array}{cc}
	    s_1 & s_2 \\ s_2^* & -s_1^*
	  \end{array}
\right)\,,
$$
which completes the proof of the theorem.

\end{IEEEproof}

\section{Three-Rate Rateless}
\label{app:rateless}

We now consider the three-rate ``rateless'' problem, as defined in \secref{ss:rateless}, with $M=L=3$ and a given rate $R$:
\begin{eqnarray*}
\HHH_1 & = & \left( \begin{array}{ccc}
	  \alpha_1 & 0 & 0 \\
	  0 &  0  & 0 \\
	  0 & 0 & 0 \\
       \end{array}
\right)
\\
\HHH_2 & = & \left( \begin{array}{ccc}
	  \alpha_2 & 0 & 0 \\
	  0 &  \alpha_2  & 0 \\
	  0 & 0 & 0
       \end{array}
\right)
\\
\HHH_3 & = & \left( \begin{array}{ccc}
	  \alpha_3 & 0 & 0 \\
	  0 &  \alpha_3  & 0 \\
	  0 & 0 &  \alpha_3
       \end{array}
\right)\,,
\end{eqnarray*}
where $\alpha_1,\alpha_2,\alpha_3$ are the positive values satisfying $\log(1 + \alpha_1^2) = 2\log(1 + \alpha_2^2) = 3 \log(1 + \alpha_3^2) = R$. As in the 2-rate case, the  covariance matrix in this problem is the identity matrix, $\CC=I$. Since $\HHH_3$ is a s scaled identity matrix, we can ignore it and concentrate on the remaining two matrices.

The augmented matrices, as defined in  \eqref{eq:G_matrix}, are:
\begin{eqnarray*}
\GGG_1 & = & \left( \begin{array}{ccc}
	  2^\frac{R}{2} & 0 & 0 \\
	  0 &  1  & 0 \\
	  0 & 0 & 1
       \end{array}
\right)
\\
\GGG_2 & = & \left( \begin{array}{ccc}
	  2^\frac{R}{4} & 0 & 0 \\
	  0 &  2^\frac{R}{4}  & 0 \\
	  0 & 0 & 1
       \end{array}
\right)\,.
\end{eqnarray*}
The decomposition \eqref{eq:jet_of_augmented} becomes:
\begin{eqnarray*}
\RRR_1 & = & 2^\frac{R}{2} \cdot \left( \begin{array}{ccc}	  
	  1 & z & w \\
	  0 &  2^{-{\frac{R}{12}}}  & x \\
	  0 & 0 & 2^{\frac{R}{12}}
       \end{array}
\right)
\\
\RRR_2 & = &  2^\frac{R}{2} \cdot \left( \begin{array}{ccc}
	  1 & z & 0 \\
	  0 &  2^{-\frac{R}{12}}  & 0 \\
	  0 & 0 & 2^{\frac{R}{12}}
       \end{array}
\right)\,,
\end{eqnarray*}
where
$$
      x = - \left(  1 - 2^{-\frac{R}{6}}   \right) \sqrt{  1 + 2^\frac{R}{6} + 2^\frac{R}{3}      }.
$$
It then follows from \thrmref{thm:perfect_two_on_two} that there exists a perfect solution over the complex field if and only if:
$$
	x^2 - 4 \leq 0\,,
$$
or explicitly:
$$
	2^{-\frac{R}{3}} \left(  1 + 2^\frac{R}{6}  \right)^2   \left(  1 - 3 \cdot 2^\frac{R}{6} + 2^\frac{R}{3}   \right) \leq 0\,.
$$
This condition is satisfied if and only if:
$$
R \leq 6 \log \left( \frac{3 + \sqrt{5}}{2}      \right) \approx 8.331\,,
$$
which coincides with the result that was obtained in \cite{etw}.

For rates higher than this threshold, a perfect capacity-achieving solution does not exist. However, as explained earlier, multiple channel usages can be utilized in order to approach capacity asymptotically.

\section{Nearly Optimal $2$-GMD for $n \geq 2$}
\label{app:large_n_proof}

We now bring the proof of \thrmref{thm:n_n_asymptotical} for the general case $n \geq 2$.

\begin{IEEEproof}[Proof of \thrmref{thm:n_n_asymptotical}]

Let $A_1$ and $A_2$ be the two $n \times n$ complex valued matrices, with determinants equal to $1$. As in \eqref{eq:extended_two}, we define the extended matrices,
\begin{equation*}
    \AAA_i = \left( \begin{array}{cccc} A_i & 0 & \cdots & 0 \\ 0 & A_i & \cdots & 0 \\ \vdots & \vdots & \ddots & \vdots \\
			0 & 0 & \cdots & A_i
                   \end{array} \right).
\end{equation*}
We are looking for three $nN \times n(N-n+1)$ matrices $\UUU_1,\UUU_2,\VVV$ with orthonormal columns, such that
$$
\UUU_i^\dagger \AAA_i \VVV
$$
are upper-triangular, with only $1$s on the diagonal.

We accomplish that using three steps:
\paragraph{Joint Triangularization}
As in the $n=2$ proof, we start by jointly triangularizing the matrices $A_1$ and $A_2$:
$$
  \Ujetidagger A_i \Vjet =   \left( \begin{array}{cccc} \smallr_1 & * & \cdots & * \\ 0 & \smallr_2 & \cdots & * \\ \vdots & \vdots & \ddots & \vdots \\
			0 & 0 & \cdots & \smallr_n
                   \end{array} \right)\,.
$$
We now apply this transformation to each block separately:
\begin{equation} \label{eq:definition_of_ggg}
		\AAA_i^\JET =   \UUUjetidagger \AAA_i \VVVjet \,,
\end{equation}
where
\begin{eqnarray*}
  \UUUjeti & = &
    \left( \begin{array}{cccc} \Ujeti & 0 & \cdots & 0 \\ 0 & \Ujeti & \cdots & 0 \\ \vdots & \vdots & \ddots & \vdots \\
			0 & 0 & \cdots & \Ujeti
                   \end{array} \right) \\
  \VVVjet & = &
    \left( \begin{array}{cccc} \Vjet & 0 & \cdots & 0 \\ 0 & \Vjet & \cdots & 0 \\ \vdots & \vdots & \ddots & \vdots \\
			0 & 0 & \cdots & \Vjet
                   \end{array} \right)\,.
\end{eqnarray*}

\paragraph{Reordering}
It will now be convenient to re-order the columns of $\AAA_i^\JET$ such that the following columns:
$$
kn,kn+(n-1),kn+2(n-1), \cdots, kn+(n-1)^2
$$
will become ``grouped together'' for every $k$.\footnote{Note that this set includes exactly one symbol from each of $n$ consecutive channel uses.} Formally, We do so by introducing the following $nN \times n(N-n+1)$ re-ordering matrix~$\OOO$:
\begin{equation} \label{eq:definition_of_O}
\OOO_{ij} =   \left\{ \begin{array}{cl}
                           1 & i=\pi_j \\
			  0 & \textrm{Otherwise,}
                      \end{array}
\right.
\end{equation}
where the function $\pi$ is defined as follows:
\begin{itemize}
 \item  For $1 \leq j \leq n(N-n+1) $,
$$
      \pi_j = (n-1) \cdot \left[ (j-1)\%n \right] + n \cdot \left\lfloor \frac{j-1}{n} \right\rfloor + n.
$$

\end{itemize}

As a result of this re-ordering, we obtain an upper-triangular $(N-n+1)n \times (N-n+1)n$ matrix, which has a block-triangular structure:
$$
\OOO^T \AAA_i^\JET \OOO = \left(
	\begin{array}{c:c:c:c:c}
	      \Lambda & * & \cdots & * & * \\ \hdashline
	      0 & \Lambda & \cdots & * & * \\ \hdashline	
              \vdots &  \vdots &   \vdots        &    \ddots    &  \vdots \\ \hdashline
	      0 & 0        & \cdots      & \Lambda & * \\              \hdashline
		0 & 0 & \cdots         &0     & \Lambda
	\end{array}
\right),
$$
where
$$
      \Lambda = \left(
      \begin{array}{cccc}
	\smallr_n & 0 & \cdots & 0 \\
	0 & \smallr_{n-1} & \cdots & 0 \\
      \vdots & \vdots & \ddots & \vdots \\
	  0 & 0 & \cdots & \smallr_1 \\
      \end{array} \right)\,.
$$

\paragraph{GMD}
Since the matrix $\Lambda$ does not depend on $i$, we can decompose it using GMD:
$$
      \Lambda = \Ugmd \bigT \Vgmddagger,
$$
where $\bigT$ is upper-triangular with only $1$s on its diagonal.

We now apply the GMD to all the blocks of $\OOO^T \AAA_i^{\JET} \OOO$:
\begin{eqnarray*}
  \UUUgmd & = &
    \left( \begin{array}{cccc} \Ugmd & 0 & \cdots & 0 \\ 0 & \Ugmd & \cdots & 0 \\ \vdots & \vdots & \ddots & \vdots \\
			0 & 0 & \cdots & \Ugmd
                   \end{array} \right) \\
  \VVVgmd & = &
    \left( \begin{array}{cccc} \Vgmd & 0 & \cdots & 0 \\ 0 & \Vgmd & \cdots & 0 \\ \vdots & \vdots & \ddots & \vdots \\
			0 & 0 & \cdots & \Vgmd
                   \end{array} \right)\,,
\end{eqnarray*}
to obtain:
\begin{equation}  \label{eq:big_btb_n}
      \UUUgmdidagger \OOO^T \GGG_i^\JET \OOO \VVVgmd =  \left(
\begin{array}{c:c:c:c:c}
\bigT_i & * & \cdots & * & * \\
 \hdashline
	      0 & \bigT_{i} & \cdots & * & * \\ \hdashline	
              \vdots &  \vdots &   \vdots        &    \ddots    &  \vdots \\ \hdashline
	      0 & 0        & \cdots      & \bigT_i & * \\              \hdashline
		0 & 0 & \cdots         &0     & \bigT_i
	\end{array} \right),
\end{equation}
where $\bigT_{i}$ are upper-triangular with only $1$s on the diagonal.

We conclude by combining \eqref{eq:big_btb_n} with \eqref{eq:definition_of_ggg} to obtain:
\begin{equation} \nonumber
      \UUU_i^\dagger \AAA_i \VVV =   \left(
\begin{array}{c:c:c:c:c}
\bigT_i & * & \cdots & * & * \\
 \hdashline
	      0 & \bigT_{i} & \cdots & * & * \\ \hdashline	
              \vdots &  \vdots &   \vdots        &    \ddots    &  \vdots \\ \hdashline
	      0 & 0        & \cdots      & \bigT_i & * \\              \hdashline
		0 & 0 & \cdots         &0     & \bigT_i
	\end{array} \right),
\end{equation}
where
\begin{eqnarray} 
\label{eq:def_of_UUU}   \UUU_i & = & \UUUjeti \OOO \UUUgmd \\
\label{eq:def_of_VVV}   \VVV_{\phantom{i}} & = & \VVVjet \OOO \VVVgmd \,,
\end{eqnarray}
which completes the proof of the theorem.

\end{IEEEproof}

\bibliographystyle{IEEEtran}
\bibliography{IEEEabrv,ISIT2011}

\begin{thebibliography}{10}
\providecommand{\url}[1]{#1}
\csname url@samestyle\endcsname
\providecommand{\newblock}{\relax}
\providecommand{\bibinfo}[2]{#2}
\providecommand{\BIBentrySTDinterwordspacing}{\spaceskip=0pt\relax}
\providecommand{\BIBentryALTinterwordstretchfactor}{4}
\providecommand{\BIBentryALTinterwordspacing}{\spaceskip=\fontdimen2\font plus
\BIBentryALTinterwordstretchfactor\fontdimen3\font minus
  \fontdimen4\font\relax}
\providecommand{\BIBforeignlanguage}[2]{{%
\expandafter\ifx\csname l@#1\endcsname\relax
\typeout{** WARNING: IEEEtran.bst: No hyphenation pattern has been}%
\typeout{** loaded for the language `#1'. Using the pattern for}%
\typeout{** the default language instead.}%
\else
\language=\csname l@#1\endcsname
\fi
#2}}
\providecommand{\BIBdecl}{\relax}
\BIBdecl

\bibitem{Telatar99}
E.~Telatar, ``Capacity of the multiple antenna {G}aussian channel,''
  \emph{Europ. Trans. Telecommun.}, vol.~10, pp. 585--595, Nov. 1999.

\bibitem{gore_book}
A.~J. Paulraj, D.~A. Gore, R.~U. Nabar, and H.~Bolcseki, ``An overview of
  {MIMO} communications - a key to gigabit wireless,'' \emph{Proceedings of the
  IEEE}, vol.~92, no.~2, pp. 198 -- 218, Feb. 2004.

\bibitem{tse_book}
D.~Tse and P.~Viswanath, \emph{Fundamentals of Wireless Communications}.\hskip
  1em plus 0.5em minus 0.4em\relax Cambridge University Press, 2005.

\bibitem{JET:SP}
\BIBentryALTinterwordspacing
A.~Khina, Y.~Kochman, and U.~Erez, ``Joint unitary triangularization for {MIMO}
  networks,'' \emph{IEEE Trans. Signal Processing}, Submitted, Dec.~2010.
  [Online]. Available: \url{http://arxiv.org/abs/1012.4715}
\BIBentrySTDinterwordspacing

\bibitem{costa}
M.~H.~M. Costa, ``Writing on dirty paper,'' \emph{IEEE Trans. Information
  Theory}, vol. IT-29, pp. 439--441, May 1983.

\bibitem{two_way_relay}
A.~Khina, Y.~Kochman, and U.~Erez, ``Physical-layer {MIMO} relaying,'' in
  \emph{Proceedings of ISIT-2011, St. Petersburg, Russia}, July 2011.

\bibitem{Wolniansky_V-BLAST}
P.~W. Wolniansky, G.~J. Foschini, G.~D. Golden, and R.~A. Valenzuela,
  ``{V-BLAST}: an architecture for realizing very high data rates over the
  rich-scattering wireless channel,'' in \emph{ISSSE 1998, URSI International
  Symposium}, pp. 295--300.

\bibitem{relay_mimo_itw}
A.~Khina, Y.~Kochman, U.~Erez, and G.~W. Wornell, ``Incremental coding over
  {MIMO} channels,'' submitted to Information Theory Workshop (ITW2011),
  Parati, Brazil.

\bibitem{alamouti}
S.~Alamouti, ``A simple transmit diversity technique for wireless
  communications,'' \emph{Selected Areas in Communications, IEEE Journal on},
  vol.~16, no.~8, pp. 1451 --1458, Oct. 1998.

\bibitem{stc}
V.~Tarokh, N.~Sheshadri, and A.~R. Calderbank, ``Space-time block codes form
  orthogonal designs,'' \emph{IEEE Trans. Information Theory}, vol. IT-45, pp.
  1456--1467, July 1999.

\bibitem{GMD}
Y.~Jiang, W.~Hager, and J.~Li, ``The geometric mean decompostion,'' \emph{Lin.
  Algebra and Its Applications}, vol. vol. 396, pp. 373--384, Feb. 2005.

\bibitem{BlackwellBreimanThomasian59}
D.~Blackwell, L.~Breiman, and A.~J. Thomasian, ``The capacity of a class of
  channels,'' \emph{The Annals of Mathematical Statistics}, vol.~30, pp.
  1229--1241, Dec. 1959.

\bibitem{etw}
\BIBentryALTinterwordspacing
U.~Erez, M.~D. Trott, and G.~W. Wornell, ``Rateless coding for {G}aussian
  channels,'' 2007. [Online]. Available: \url{http://arxiv.org/abs/0708.2575}
\BIBentrySTDinterwordspacing

\bibitem{orthogonal_design}
A.~V. Geramita and J.~Seberry, ``Orthogonal designs, quadratic forms and
  {H}adamard matrices,'' \emph{Lec.~Notes in Pure and App.~Mathematics}, 1979.

\end{thebibliography}

\end{document}